\documentclass[amsmath,amssymb,aps,twocolumn,superscriptaddress,prb,10pt]{revtex4-2}

\usepackage{graphicx}
\graphicspath{{figures/}}
\usepackage{hyperref}
\usepackage{MnSymbol}
\usepackage[normalem]{ulem}
\usepackage{float}

\DeclareMathOperator{\Tr}{Tr}

\DeclareMathOperator{\const}{const}

\begin{document}

\title{Quantum dynamics of monitored free fermions: Evolution of quantum correlations and scaling at measurement-induced phase transition}

\author{Igor Poboiko}
\author{Alexander D. Mirlin}
\affiliation{\mbox{Institute for Quantum Materials and Technologies, Karlsruhe Institute of Technology, 76131 Karlsruhe, Germany}}
\affiliation{\mbox{Institut f\"ur Theorie der Kondensierten Materie, Karlsruhe Institute of Technology, 76131 Karlsruhe, Germany}}
\date{\today}

\begin{abstract}

We explore, both analytically and numerically, the quantum dynamics of a many-body free-fermion system subjected to local density measurements. We begin by extending the mapping to the nonlinear sigma-model (NLSM) field theory for the case of finite evolution time $T$ and different classes of initial states, which lead to different NLSM boundary conditions. The analytical formalism is then used to study how quantum correlations gradually develop, with increasing $T$, from those determined by the initial state towards their steady-state form. The analytical results are confirmed by numerical simulations for several types of initial states. We further consider the long-time limit, when the system in $d+1$ space-time dimensions becomes quasi-one-dimensional, and analyze the scaling of the ``localization'' time (which is simultaneously the purification time and the charge-sharpening time for this class of problems). The analytical predictions for scaling properties are fully confirmed by 
numerical simulations in a $d=2$ model around the measurement-induced phase transition. We use this dynamical approach to determine numerically the measurement-induced transition point and the associated correlation-length critical exponent.

\end{abstract}

\maketitle

\section{Introduction}

Quantum measurements play a very special role in quantum theory, in view of the non-unitary character of the associated dynamics of a quantum state. Specifically, a measurement induces a  stochastic wave-function collapse (strong or weak, depending on the measurement strength) that depends on the measurement outcome. 
Effects of quantum measurement on properties of many-body quantum systems are attracting much of research attention.  
It was found that, due to a competition between  quantum measurements and unitary dynamics, resulting quantum states may exhibit transitions between phases with different scaling of the entanglement entropy $S$  with the size $\ell$ of a subsystem \cite{Li2018a, Skinner2019a, Chan2019a, Cao2019a, Szyniszewski2019a,  Li2019a, Bao2020a}, see also reviews \cite{Potter2022, Fisher2022} and Refs.~\cite{Noel2022a,Koh2022,Hoke2023,Agrawal2023} for experimental studies. 
For frequent measurements, the entanglement entropy scales as the area of the subsystem boundary $\sim \ell^{d-1}$ (area-law phase), implying that the entanglement is localized in the boundary region. With lowering measurement rate, a transition into a phase with a faster increase of $S(\ell)$ may take place, characterized by delocalization of quantum information.

An important class of monitored many-body systems is formed by free-fermion systems, with measurements preserving the Gaussianity of a quantum state (i.e., its Slater-determinant character for pure states)  \cite{Cao2019a, Turkeshi2021, Alberton2021a, Buchhold2021a, Coppola2022, Carollo2022, 
Szyniszewski2022, Jian2023, Fava2023, Poboiko2023a, Poboiko2023b, chahine2023entanglement, Lumia2023, starchl2024generalized, FavaNahum2024}. 
For the case of complex free fermions, it was shown that, in $d=1$ spatial dimension,  the system is  in the area-law phase for any non-zero monitoring rate
(in the asymptotic limit $\ell \to \infty$)
 \cite{Poboiko2023a, FavaNahum2024}. For higher spatial dimensionality, $d > 1$, the system exhibits a phase transition between the area-law phase and a phase with  a long-range entanglement characterized by $\ell^{d-1} \ln \ell$ scaling of $S(\ell)$ 
\cite{Poboiko2023b, chahine2023entanglement}. These analytical results have been obtained by mapping to a replica non-linear sigma model (NLSM), supplemented by its renormalization-group analysis. 
They were also confirmed and complemented by numerical simulations.
In Refs.  \cite{Poboiko2023a,Poboiko2023b} an analogy between the measurement-induced phase transition (MIPT) in $d$ dimensions and Anderson-localization transition in $d+1$ dimensions was emphasized. The two problems are described by closely related NLSM field theories, although with important differences in the replica limit and the symmetry class. 
We also refer the reader to Refs.~\cite{Jian2022,Jian2023} for a broad discussion of field theories of random non-unitary dynamics and comparison with theories of Anderson localization.

The results quoted above concern the steady ensemble of quantum states that emerges when the combination of unitary dynamics and quantum measurements acts on a system for a very long time, $T \to \infty$. In this situation, properties of the resulting quantum states do not depend on the initial state. At the same time, one can ask a question of quantum dynamics of a monitored system at $T<\infty$.
 Some aspects of this problem were discussed in the MIPT context for particular models and initial states \cite{Fava2023,FavaNahum2024,XiaoKawabata2024,Mochizuki2025,HisanoriFuji2025,Pan2025,Xiao2025universal}. It has been understood that dynamical properties allow one to define the ``purification'' and ``charge-sharpening'' phase transitions for random quantum circuits subject to measurements 
\cite{Gullans2020a,Zabalo2020a,Gullans2020scalable,Bentsen2021measurement,
Lunt2021measurement-induced,Agrawal2022,Loio2023}.

Our goal here is to explore, both analytically and numerically, the dynamical evolution of a monitored many-body free-fermion system. One of the key questions that we address is: How do quantum correlations evolve from the initial state towards this asymptotic ensemble as the evolution time $T$ increases, from relatively short to long times? Clearly, the answer will in general depend not only on the system under consideration but also on the type of initial state, and we also analyze this dependence. A related set of questions concerns the dynamical scaling (at large but finite $T$) around the MIPT in a system of complex free fermions \cite{Poboiko2023b} studied earlier by considering the $T\to \infty$ limit. Can this scaling be efficiently used to determine critical exponents for this transition? If yes, how do the results compare to those obtained from studying the $T\to\infty$ system? A success here would pave the way to employing the quantum dynamics for efficient investigation of critical behavior at free-fermion MIPTs of other universality classes.

We begin by extending
in Sec.~\ref{sec:initial_conditions}
the mapping to NLSM field-theory   \cite{Poboiko2023a,Poboiko2023b} for the case of finite evolution time $T$ and different initial states. Specifically, we consider three classes of Gaussian initial states---maximally mixed state (which also exhibits maximal charge fluctuations out of all Gaussian states), random area-law pure state, and random volume-law pure state---and show which boundary conditions for the NLSM correspond to each of them. We then use this general formalism in Sec.~\ref{sec:diffusive} to study the time dependence of quantum correlations at relatively short times $T$
for different initial conditions. The results, demonstrating a gradual development of $T\to \infty$ quantum correlations with increasing $T$, are supported by numerical simulations of a $d=1$ system for all three types of initial states. Last but not least, we use in Sec.~\ref{sec:MIPT} the long-time dynamics to get a complementary view on  the MIPT in a $d=2$ system and to study it numerically. Scaling properties of the quasi-one-dimensional ``localization length'' in time direction (the inverse Lyapunov exponent) are in full agreement with 
analytical expectations. Further, the numerical values of the MIPT critical measurement rate and the localization-length critical exponent are consistent with those originally obtained in Ref.~\cite{Poboiko2023b} 
(see also a more recent work \cite{fan2025entanglement})
by studying scaling properties of $T\to \infty$ states.
Our work opens up broad prospects for investigation of quantum dynamics and MIPTs in monitored many-body systems, 
as discussed in Sec.~\ref{sec:summary}.

\section{Field theory and initial conditions}
\label{sec:initial_conditions}

We consider a system of non-interacting complex fermions on a $d$-dimensional lattice evolving under a time-independent local Hamiltonian $H$ and local density measurements performed randomly in space and time with a rate $\gamma$. The simplest model is that considered in Refs.~\cite{Poboiko2023a,Poboiko2023b} with $H$ given by 
\begin{equation}
H=-J\sum_{\left\langle \boldsymbol{x} \boldsymbol{x}^\prime\right\rangle }\left(\psi_{\boldsymbol{x}}^{\dagger}\psi_{\boldsymbol{x}^\prime}+\text{H.c.}\right),
\label{eq:H}
\end{equation}
where the sum goes over nearest-neighbor site pairs,
and with projective measurements of site particle numbers $n_{\boldsymbol{x}} = \psi_{\boldsymbol{x}}^\dagger \psi_{\boldsymbol{x}}$. We will use this model in our numerical simulations below. We emphasize, however, that the analytical part of this work is much more general. Specifically, it applies to any microscopic model of the same universality class, including models with weak measurements considered, e.g., in Refs.~\cite{Cao2019a,chahine2023entanglement,FavaNahum2024,fan2025entanglement}. Indeed, these models can be mapped onto the same NLSM field theory, implying the same physics (for spatial and temporal scales above the mean free path $\ell_0 \sim J / \gamma$ and mean free time $\sim 1/\gamma$, respectively). 

The classification of monitored free fermions is based on the symmetries of the Keldysh fermionic field theory description. Such description has an inherent chiral symmetry owing to the two chiral branches of the Keldysh contour.
Thus, in the absence of additional symmetries, the system would belong to the chiral unitary class AIII. 
An equivalent way to identify the symmetry class is based on NLSM target space, which in this case is $\mathrm{SU}(R)$, which corresponds, up to a different replica limit, to a disordered system whose Hamiltonian belongs to the class AIII. This class will be assumed for simplicity in our analytical treatment.
In fact, the system with nearest-neighbor Hamiltonian \eqref{eq:H} and site charge monitoring, which we will study numerically below, has an additional effective particle-hole symmetry and belongs to the chiral orthogonal symmetry class BDI, see Refs.~\cite{FavaNahum2024,Poboiko2025}.
This particle-hole symmetry can be broken, e.g., by adding a next-nearest-neighbor hopping in Eq.~\eqref{eq:H}, reducing the symmetry to that of class AIII. 
The difference between these two classes is minor in the context of this work and will not play any role for our analytical considerations below. 
Specifically, the difference does not manifest itself in the diffusive regime of relatively short times $T$ studied analytically in Sec.~\ref{sec:diffusive}, where quantum corrections are parametrically small and can be neglected. 
Furthermore, the overall structure of quantum corrections in the two classes is very similar, and both classes support a phase transition for spatial dimensions $d > 1$ in the replica limit $R \to 1$. 
Further, scaling properties of the quasi-one-dimensional ``localization length'' $T^*$ around the MIPT, discussed in Sec.~\ref{sec:MIPT}, are fully analogous in both classes.
We note, however, that values of the critical exponents and exact form of the scaling functions are generally expected to depend on the symmetry class.

The NLSM describes Goldstone modes associated with continuous symmetry characterizing a Gaussian system, that is unitary transformations of $R$ replicas of the fermionic fields in the Keldysh path integral formalism. The replica trick is utilized to perform averaging over different quantum trajectories, with the replica limit $R \to 1$ incorporating the Born's rule for different measurement outcomes. As result, the action of the NLSM in $(d+1)$-dimensional space-time reads (we follow the notations of Refs.~\cite{Poboiko2023a,Poboiko2023b,Poboiko2025}):
\begin{equation}
\label{eq:NLSM}
S[\hat{U}]=\frac{g}{2}\int d^{d}\boldsymbol{x}\, dt\Tr\left(v^{-1} \partial_{t}\hat{U}^{\dagger}\partial_{t}\hat{U}+v \nabla\hat{U}^{\dagger}\nabla\hat{U}\right),
\end{equation}
where $\hat{U}(\boldsymbol{x},t) \in \mathrm{SU}(R)$ is a matrix field corresponding to the Goldstone modes described above, $g \propto J / \gamma$ is the coupling constant akin to the diffusion constant in the context of disordered systems, and $v \propto J$ is the characteristic velocity. Note that this action acquires isotropic form in space-time in the coordinates $(\boldsymbol{x}, v t)$.

The central quantity characterizing a Gaussian quantum state for a given quantum trajectory is the two-point density correlation function
\begin{equation}
    \mathcal{C}_{\boldsymbol{x} \boldsymbol{x}^\prime}=\left\langle n_{\boldsymbol{x}}n_{\boldsymbol{x}^\prime}\right\rangle -\left\langle n_{\boldsymbol{x}}\right\rangle \left\langle n_{\boldsymbol{x}^\prime}\right\rangle.
\end{equation}
The key observables that characterize quantum correlations can be expressed in terms of $\mathcal{C}_{\boldsymbol{x}\boldsymbol{x}^\prime}$. 
In particular, the variance of charge fluctuation in a region $A$ is given by
\begin{equation}
\label{eq:CA2}
\mathcal{C}_{A}^{(2)}=\left\langle \left\langle \hat{N}_{A}^{2}\right\rangle \right\rangle =\sum_{\boldsymbol{x},\boldsymbol{x}^{\prime}\in A}\mathcal{C}_{\boldsymbol{x}\boldsymbol{x}^{\prime}}.
\end{equation}
The entropy of the region $A$,
\begin{equation}
S_{A} = -\Tr(\rho_A \ln \rho_A).
\end{equation}
is related (for a Gaussian quantum state) to the charge variance $\mathcal{C}_{A}^{(2)}$ via
\begin{equation}
S_{A} \approx\frac{\pi^{2}}{3}\mathcal{C}_{A}^{(2)},
\label{eq:S-C}
\end{equation}
which is a truncated version of the Klich-Levitov relation, Ref.~\cite{KlichLevitov}. While the exact formula also involves higher-order charge cumulants $\mathcal{C}_A^{(2q)}$ with $q > 1$, they were shown to be parametrically small for $\gamma \ll J$ and numerically small for $\gamma \gtrsim J$, see Ref.~\cite{Poboiko2023a}.
When the considered system is in a pure state, $S_A$ is the entanglement entropy between A and the rest of the system. Further, the particle-number covariance between two regions $A$ and $B$ is expressed in terms of $\mathcal{C}_{\boldsymbol{x}\boldsymbol{x}^\prime}$ as
\begin{equation}
G_{AB}=\left\langle N_{A}\right\rangle \left\langle N_{B}\right\rangle -\left\langle N_{A}N_{B}\right\rangle =-\sum_{\boldsymbol{x}\in A,\boldsymbol{x}^{\prime}\in B}\mathcal{C}_{\boldsymbol{x}\boldsymbol{x}^{\prime}}.
\end{equation}
For two non-intersecting regions A and B, the covariance $G_{AB}$ is related to the mutual information $\mathcal{I}(A:B)$ via
\begin{equation}
\mathcal{I}(A:B)\approx\frac{2\pi^{2}}{3}G_{AB}.
\end{equation}

In terms of the NLSM, the density correlation function averaged over different quantum trajectories is expressed as a correlation function of NLSM currents (see Appendix~\ref{sec:app:NLSM:current}):
\begin{multline}
\label{eq:C}
\mathcal{C}(\boldsymbol{x},\boldsymbol{x}^{\prime})\equiv\overline{\mathcal{C}_{\boldsymbol{x}\boldsymbol{x}^{\prime}}}\\=\lim_{t,t^{\prime}\to0}\left[\frac{g}{v}\delta(\boldsymbol{x}-\boldsymbol{x}^{\prime})\delta(t-t^{\prime})-\frac{g^2}{v^2}\left\langle {\cal J}_{ab}(\boldsymbol{x},t){\cal J}_{ba}(\boldsymbol{x}^{\prime},t^{\prime})\right\rangle \right],
\end{multline}
where $a \neq b$ denote two arbitrary different replicas, and $\hat{{\cal J}}=-i \hat{U}^{\dagger}\partial_{t}\hat{U}$ denotes the temporal component of the Noether current. 
The points $\boldsymbol{x}$ and $\boldsymbol{x}^\prime$ belong to the $d$-dimensional surface of the space-time that corresponds to the time (that we choose to be $t=0$) at which the observables are studied. The NLSM is supplemented with absorbing boundary conditions at this surface (which follows from the fact that measurements at times $t>0$ cannot affect properties of the system at $t=0$). This expression is exact within the NLSM theory, and describes the long-wavelength limit $q \ell_0 \ll 1$ of the original microscopic problem.

As emphasized in Refs.~\cite{Jian2023, Poboiko2023a,Poboiko2023b}, the problem has close similarity to the Anderson localization problem in $(d+1)$ dimensions, with random measurement positions and outcomes playing a role of ``impurities'' in space-time. Important differences include the replica limit ($R\to 1$ here instead of $R\to 0$ for the Anderson-localization problem) and the symmetry class (which is BDI or AIII for the models under consideration, as discussed above), with the chiral symmetry emerging due to the structure of the action in the Keldysh space. 
Since $\mathcal{C}(\boldsymbol{x},\boldsymbol{x}^\prime)$ is given by the current-current correlation function of the NLSM, it is a counterpart of the two-point conductance of the Anderson-localization problem. Correspondingly, the particle-number covariance $G_{AB}$ is a counterpart of the two-terminal conductance. 

\begin{figure}
    \centering
   \includegraphics[width=\columnwidth]{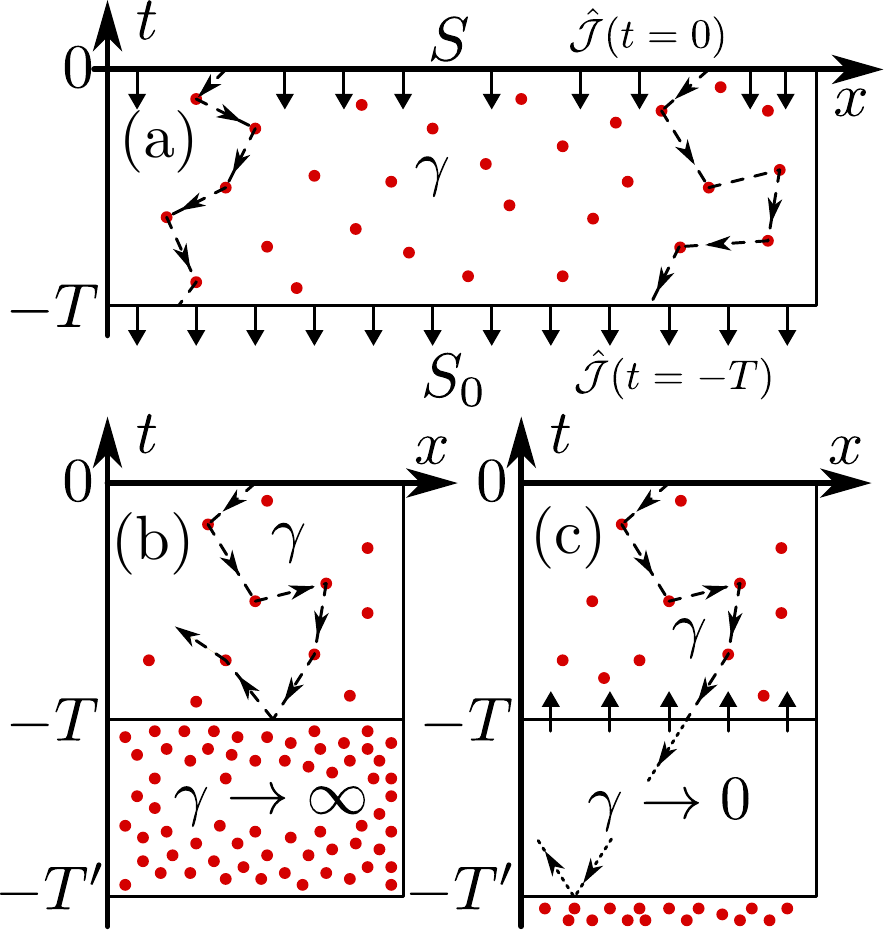}
    \caption{Boundary conditions for NLSM field theory: Schematic space-time representation of a monitored evolution subject to different initial conditions. 
    Red dots represent individual measurements. The system is evolved with some measurement rate $\gamma$ from the initial time $t=-T$ to time $t=0$ where the correlation function
    $\mathcal{C}_{\boldsymbol{x} \boldsymbol{x}^\prime}$  and related
    observables are studied. The boundary condition at the $t=0$ boundary is always absorbing.
    (a) Initial state is a maximally mixed state.
    Due to global current conservation $\hat{\mathcal{J}}(t=-T) = \hat{\mathcal{J}}(t=0)$, schematically represented by arrows, this corresponds to maximal ``conductance'' between $S$ and $S_0$, which is realized with the absorbing boundary condition for the NLSM at $t = -T$, Eq.~\eqref{eq:BC:absorbing}. (b) Initial state a maximally disentangled (random bitstring) state. It can be realized from an arbitrary random state at $t=-T^\prime < -T$ by evolving it to $t = -T$ with an infinite measurement rate $\gamma \to \infty$. This corresponds to an ``insulating'' system for $t \in [-T^\prime,-T]$; thus the boundary condition at $t = -T$ is reflecting, Eq.~\eqref{eq:BC:reflecting}. (c) A volume-law pure state can be realized from random bitstring at $t = -T^\prime \to -\infty$ (corresponding to the reflecting boundary, as in (b)) by evolving it unitarily ($\gamma \to 0$) up to time $t = -T$. This corresponds to a perfect conductor at $t \in [-T^\prime,-T]$, so that the boundary condition at $t=-T$ is absorbing. However, all the current that flows down through this boundary eventually gets reflected at $t = -T^\prime$ and returns back to the system homogeneously through the whole $t = -T$ surface. Thus, owing to the total current conservation, the $\boldsymbol{q} = 0$ mode gets reflected, see Eq.~\eqref{eq:BC:mixed}.}
    \label{fig:boundary}
\end{figure}

These results were derived in Refs.~\cite{Poboiko2023a,Poboiko2023b} where the limit of long time evolution ($T\to \infty$) from the initial state at $t_i=-T$ was considered. In this limit, a steady ensemble of quantum states is formed, which does not depend on the initial state. Clearly, for finite $T$ some information about the initial state should remain, so that the NLSM theory should be supplemented by a boundary condition at $t=t_i$ that depends on the initial state. We will now consider three important classes of initial states and determine the corresponding $t=t_i$ boundary conditions for the NLSM, which are illustrated  in Fig.~\ref{fig:boundary}.

\paragraph{Maximally mixed state.}

First, we consider the case when the initial state is a maximally mixed state characterized by infinite-temperature density matrix
\begin{equation}
\hat{\rho}=2^{-L^{d}}\,\hat{\mathbb{I}}\,.
\end{equation}
To understand the corresponding NLSM boundary condition, we note that, out of all Gaussian states, this state provides the maximum variance of the total particle number (charge). In terms of the 
NLSM, the charge variance in the final state (i.e., the one at time $t=0$) is given by minus the conductance $- G_{S,S}$, where $G_{S,S}$ is the ``self-conductance'' and $S$ is the whole $t=0$ surface. In view of current conservation, the total current entering the system through $S$ should be equal to the total current leaving the system through the $t=t_i$ surface $S_0$. Thus, the charge variance at time $t=0$ (i.e., after the evolution over the time $T$) is equal to $G_{S,S_0}$. Therefore, the NLSM boundary condition corresponding to the maximally mixed state should provide the maximal value of $G_{S,S_0}$. Clearly, this is the fully absorbing boundary condition:
\begin{equation}
\label{eq:BC:absorbing}
\hat{U}(\boldsymbol{x},t_{i})=\hat{\mathbb{I}}\,.
\end{equation}
On a more formal level, this initial condition for a maximally mixed state can be derived from the fermionic Keldysh path integral representation, see Appendix~\ref{sec:app:boundary}. It was also derived for a different microscopic model in the same universality class in Ref.~\cite{FavaNahum2024} (see also an analogous derivation for a model of monitored Majorana fermions in Ref.~\cite{Fava2023}).

\paragraph{Maximally disentangled state.}
 We consider now the case when the initial state is a maximally disentangled (area-law) random pure state, i.e., a random bitstring on the lattice. To understand the associated boundary condition, we notice that such a state can be obtained starting from any initial state at an earlier time $t=-T' < -T$ by applying the measurement protocol with the measurement rate $\gamma' \to \infty$. In terms of the NLSM, this will imply the diffusion constant $g^\prime \to 0$ in the time interval $[-T', -T]$. The corresponding part of the space-time sample will be therefore an insulator. Thus, the NLSM boundary condition at $t=-T$ should be that for a boundary with an insulating region. This is the reflecting boundary condition:
\begin{equation}
\label{eq:BC:reflecting}
\hat{{\cal J}}(\boldsymbol{x},t_{i})=0 \,.
\end{equation}

\paragraph{Volume-law pure state.}
Finally, we consider the case of an initial state being a random volume-law pure state. (A good representative of this class of states is a random bitstring in the momentum space.) To understand the NLSM boundary condition in this situation, we construct such a state in the following way. Let us start with an initial state in the form of a random area-law state (discussed above) at an earlier time $t=-T' < -T$, with a sufficiently large time interval $T' - T$. Further, let us assume the measurement rate $\gamma' \to 0$ in the time interval $[-T',-T]$. This will generate the required random volume-law pure state at $t=-T$. In terms of the NLSM, the system in the range $[-T',-T]$ will be characterized by a diffusion constant $g^\prime \to \infty$, i.e., this part of the space-time sample will be a ``perfect metal''. This implies an absorbing boundary condition at $t=-T$.  However, there is one correction to this conclusion. Indeed, the total current is reflected at the $t=-T'$ boundary, and thus, it should also be reflected at the $t=-T$ boundary. 
Therefore, the NLSM boundary condition at $t=-T$ in this case is absorbing for all modes with non-zero wave-vectors, $q\ne 0$, and reflecting for the zero mode $q=0$. This corresponds to the boundary condition
\begin{equation}
\label{eq:BC:mixed}
\hat{U}(\boldsymbol{x},t_{i})=\const,\quad \int d^{d}\boldsymbol{x}\:\hat{\mathcal{J}}(\boldsymbol{x},t_i)=0 \,,
\end{equation}
i.e., an arbitrary spatially homogeneous matrix field configuration with a zero total current through the boundary.

\section{Dynamics of quantum correlations in the diffusive regime}
\label{sec:diffusive}

Let us now discuss relevant time scales for the dynamical problem. 
As discussed in Sec.~\ref{sec:initial_conditions}, the effect of finite time $T$ is related, in the NLSM framework, to current propagation from the $t=0$ to $t=-T$ boundary. An analogy to a $(d+1)$-dimensional disordered system then provides a physical picture of the corresponding transport regimes. For shortest times, $T \ll \ell_0 / v \sim \gamma^{-1}$, the transport in time direction is ballistic. This means that measurements simply do not have time to affect the system in any essential way, so that the system at $t=0$ has approximately the same properties as the initial state at $t=-T$. In the opposite limit of very long times $T$, we are dealing with the quasi-one-dimensional geometry of the $(d+1)$-dimensional system.
 Such a system necessarily exhibits Anderson localization in the time direction, with some characteristic time scale (localization length) $T^{\ast}$. This time scale determines the rate at which the memory of the initial state becomes exponentially lost. 
 In particular, any Gaussian state becomes pure at $T \gg T^\ast$, so that $T^\ast$ can be associated with the purification time scale. We will analyze the scaling of $T^\ast$ (which also plays the role of charge-sharpening time) in Sec.~\ref{sec:MIPT}, where we will use the long-time dynamics at 
 $T \gg T^\ast$ to study the MIPT. 
 
 In this Section, we will discuss the dynamics at sufficiently short times 
(yet larger than the mean-free time), $\ell_0 / v \ll T \ll T^\ast$, when the system is diffusive. This diffusive regime of times $T$ is parametrically broad when the measurement rate $\gamma$ is sufficiently small. 
Specifically, for $d > 1$, the condition is that $\gamma$ is below the critical value $\gamma_c$ of MIPT. For $d=1$, there is a parametrically broad diffusive regime if $\gamma \ll 1$.

\subsection{Analytics}
\label{sec:diffusive-analytics}

In the diffusive regime, we can treat the NLSM field theory \eqref{eq:NLSM} in the Gaussian approximation, $\hat{U}(\boldsymbol{x},t)=\exp\left(i\hat{\Phi}(\boldsymbol{x},t)\right)\approx\hat{\mathbb{I}}+i\hat{\Phi}(\boldsymbol{x},t)$ and $\hat{{\cal J}}(\boldsymbol{x},t)\approx \partial_{t}\hat{\Phi}(\boldsymbol{x},t)$. Within this approximation, the Fourier transform of the correlation function \eqref{eq:C} is directly expressed in terms of the Green's function of the Laplace equation:
\begin{equation}
\mathcal{C}(\boldsymbol{q},T)=\frac{g}{v}\lim_{t,t^{\prime}\to0}\left[\delta(t-t^{\prime})-\partial_{t}\partial_{t^{\prime}}G(\boldsymbol{q},t,t^{\prime})\right],
\label{eq:Cq-G}
\end{equation}
where $G$ satisfies the equation
(with $q \equiv |\boldsymbol{q}|$)
\begin{equation}
\left(v^2q^{2}-\partial_{t}^{2}\right)G(\boldsymbol{q},t,t^{\prime})=\delta(t-t^{\prime})
\end{equation}
supplemented by boundary conditions at $t=0$ and $t=-T$. Specifically, the boundary condition at $t = 0$ is absorbing, whereas the boundary condition at $t = -T$ depends on the character of the initial state, as discussed in Sec.~\ref{sec:initial_conditions}. We analyze now the implications of different  boundary conditions at $t = -T$ for the density correlation function.

\paragraph{Maximally mixed initial state.}
In this case, i.e., for the absorbing initial-time boundary condition, $\Phi(t = -T) = 0$, the Green's function reads:
\begin{equation}
G(\boldsymbol{q},t,t^{\prime})=-\frac{\sinh\left(v q\left(t_{<}+T\right)\right)\sinh\left(v q t_{>}\right)}{v q\sinh\left(v q T\right)},
\end{equation}
where $t_{<}=\min(t,t^{\prime})$ and $t_{>} = \max(t, t^\prime)$. Upon substitution in Eq.~\eqref{eq:Cq-G}, this yields
\begin{equation}
\label{eq:CqAbsorbing}
\mathcal{C}(\boldsymbol{q},T)=gq\coth\left(v q T\right).
\end{equation}
In the limit $v q T \gg 1$, Eq.~\eqref{eq:CqAbsorbing} reduces to the steady-state result \cite{Poboiko2023a} for the diffusive regime, $\mathcal{C}(\boldsymbol{q})=gq$. On the other hand, for smaller $v q T$, the correlation function has not reached its asymptotic limit. In particular, there is a finite limit $\mathcal{C}(q = 0)=g/v T$, implying the volume-law fluctuations of charge in the whole system, which decay with time as $1/T$, 
\begin{equation}
\label{eq:CT:diffusive}
{\cal C}^{(2)}(T)=L^{d}{\cal C}(\boldsymbol{q}=0,T)=\frac{gL^{d}}{v T},
\end{equation}
and an analogous behavior of the entropy, $S(T)\approx(\pi^{2}/3){\cal C}^{(2)}(T)$. It is worth reiterating that such behavior is established at time scales of the order of the mean-free time, $T\gtrsim\ell_{0} / v$, whereas at shorter times the behavior is determined by the initial state, ${\cal C}^{(2)}(0) \approx L^{d}/4$ and $S(0) \approx L^{d}\ln2$.

The result \eqref{eq:CqAbsorbing} for the correlation function allows us to calculate also the variance of charge fluctuations in a subsystem $A$ by using Eq.~\eqref{eq:CA2}. For the whole system of size $L^d$, we choose the subsystem $A$ to be of a strip geometry with dimensions $L^{d-1} \times \ell_A$, with $\ell_0 \ll \ell_A \ll L$.
We find then, according to Eq.~\eqref{eq:CA2}, 
\begin{equation}
{\cal C}_{A}^{(2)}(T)
= \frac{2}{\pi}L^{d-1}\int_{0}^{\infty}dq_{\perp}{\cal C}(q_{\perp},T)\frac{1-\cos(q_{\perp}\ell_A)}{q_{\perp}^{2}},
\label{eq:CA2-T-strip-general}
\end{equation}
where $q_\perp$ is the momentum component perpendicular to the strip (i.e., along the $\ell_A$ direction). Calculating the integral, we find
\begin{multline}
{\cal C}_{A}^{(2)}(T)=\frac{2g}{\pi}L^{d-1}\ln\left(\frac{2 v T}{\pi\ell_{0}}\sinh\frac{\pi\ell_{A}}{2 v T}\right)\\\approx\begin{cases}
g\left\Vert A\right\Vert /v T, & \ell_{A}\gg v T,\\
(g/\pi)\left\Vert \partial A\right\Vert \ln(\ell_A/\ell_{0}), & \ell_{A}\ll v T,
\end{cases}
\label{eq:CA2-T-max-mixed}
\end{multline}
where $\left\Vert A\right\Vert = L^{d-1}\ell_{A}$ is the volume of the subsystem and $\left\Vert \partial A\right\Vert =2L^{d-1}$ is its surface area. Thus, at sufficiently short times (or for sufficiently large subsystems), we have a volume-law behavior $\mathcal{C}_{A}^{(2)}\propto\left\Vert A\right\Vert$, with the prefactor decaying with time as $1/T$, whereas at larger times (or smaller subsystems) it crosses over to a time-independent steady-state area$\times$log-law behavior, $\mathcal{C}_{A}^{(2)}\propto\left\Vert \partial A\right\Vert \ln(\ell/\ell_{0})$. We note that the volume-law behavior at $\ell_A \gg v T$, Eq.~\eqref{eq:CA2-T-max-mixed}, originates from finite value $\mathcal{C}(q = 0) = g / v T$, and, in fact, holds up to subsystem sizes $\ell_A \sim L$, matching Eq.~\eqref{eq:CT:diffusive} for $\ell_A = L$.
According to Eq.~\eqref{eq:S-C}, these results hold also for the entanglement entropy $S_A$ of the subsystem (with an additional overall numerical prefactor $\pi^2/3$). 

The steady-state regime establishes in Eq.~\eqref{eq:CA2-T-max-mixed} at $v T > \ell_A$. This implies that, with increasing $T$, the quantum correlations propagate ballistically with the velocity $v$. We can also see this by calculating the real-space correlation function. 
Fourier transforming the result \eqref{eq:CqAbsorbing} and choosing the $d=1$ case for definiteness, we find the density correlation function in real space,
\begin{eqnarray}
{\cal C}(x,T)&=&-\frac{g}{\pi}\left(\frac{2 v T}{\pi}\sinh\frac{\pi x}{2 v T}\right)^{-2}
\nonumber \\
&\approx&\begin{cases}
-g/\pi x^{2}, & x\ll v T\,; \\[0.2cm]
\displaystyle -\frac{g\pi}{v^{2} T^{2}}\exp\left(-\pi x/ v T\right), & x\gg v T \,.
\label{eq:Cx-absorbing}
\end{cases}
\end{eqnarray}
This result demonstrates how the power-law correlations characteristic for the steady state gradually develop, propagating to larger and larger distances ($\sim T$) with increasing time $T$. 
It is easy to see that the same picture holds also for higher spatial dimensionalities $d$ (where the power-law correlations scale as $1/x^{d+1}$).

\paragraph{Maximally disentangled initial state.}

For this type of the initial state, i.e., for the reflecting boundary condition, $\Phi^\prime(t=-T) = 0$, the Green's function reads
\begin{equation}
G(\boldsymbol{q},t,t^{\prime})=\frac{\cosh\left(v q\left(t_{<}+ T\right)\right)\sinh\left(v qt_{>}\right)}{v q\cosh\left(v q T\right)},
\end{equation}
leading to
\begin{equation}
\label{eq:CqReflecting}
\mathcal{C}(\boldsymbol{q},T)=gq\tanh(v q T).
\end{equation}
For $v q T \gg 1$ it reduces to the same steady-sate limit, $\mathcal{C}(\boldsymbol{q})=gq$, as Eq.~\eqref{eq:CqAbsorbing}. On the other hand, for small $v qT$, it shows a very different behavior, with $\mathcal{C}(\boldsymbol{q}=0, T)=0$, which is a manifestation of the absence of fluctuations of the total charge (which is conserved under dynamics) in the initial state.

Calculating the variance of charge fluctuations in a subregion $A$ of strip geometry by using Eq.~\eqref{eq:CA2-T-strip-general}, we obtain
\begin{multline}
{\cal C}_{A}^{(2)}(T)=\frac{2g}{\pi}L^{d-1}\ln\left(\frac{4v T}{\pi\ell_{0}}\tanh\frac{\pi\ell_{A}}{4v T}\right)\\
\simeq\frac{g}{\pi}\left\Vert \partial A\right\Vert \ln\frac{\min\left(v T,\ell_{A}\right)}{\ell_{0}}.
\label{eq:CA2-disentangled}
\end{multline}
This result shows that fluctuations, which were initially absent, follow at shorter times, $v T < \ell_A$, the area law with a coefficient growing logarithmically with time, ${\cal C}_{A}^{(2)}(T)\simeq (2g/\pi)L^{d-1}\ln(v T/\ell_{0})$. Then, at times $v T \sim \ell_A$, this behavior crosses over to the steady-state area$\times$log behavior ${\cal C}_{A}^{(2)}(T)=(2g / \pi)L^{d-1}\ln(\ell / \ell_0)$. 
The entanglement entropy $S_A$ exhibits the same behavior according to Eq.~\eqref{eq:S-C}. 

The $d=1$ Fourier transform now reads
\begin{eqnarray}
\hspace*{-0.5cm} \mathcal{C}(x,T) &=& - \frac{g}{\pi} \left\{2\left[\frac{4 v T}{\pi}\sinh\left(\frac{\pi x}{4 v T}\right)\right]^{-2} \right. \nonumber \\
 & - & \left. \left[\frac{2 v T}{\pi}\sinh\left(\frac{\pi x}{2 v T}\right)\right]^{-2}\right\}
\nonumber \\
&\approx&\begin{cases}
-g/\pi x^{2}, & x\ll v T\,; \\[0.2cm]
\displaystyle 
-\frac{g\pi}{ 2 v^{2} T^{2}}\exp\left(-\pi x/ 2v T\right), & x\gg vT \,.
\label{eq:Cx-reflecting}
\end{cases}
\end{eqnarray}
The power-law correlations in real space are gradually established with increasing $T$, similarly to the case of a maximally mixed initial state, Eq.~\eqref{eq:Cx-absorbing}.

\paragraph{Volume-law pure initial state.} In this case, the initial-time boundary condition for all momenta $\boldsymbol{q}\ne 0$ is absorbing, i.e., the same as for a maximally mixed state. 
Thus, Eq.~\eqref{eq:CqAbsorbing} holds for all  $\boldsymbol{q}\ne 0$. At the same time, the reflecting boundary condition for $\boldsymbol{q}=0$ yields
\begin{equation}
G(\boldsymbol{q}=0,t,t^{\prime})=\max(t,t^{\prime}),
\end{equation}
implying ${\cal C}(\boldsymbol{q}=0,T) = 0$ instead of $\mathcal{C}(\boldsymbol{q}=0,T)=g/v T$ that we had for the maximally mixed initial state. 

Singularity at $\boldsymbol{q} = 0$ manifests itself in charge fluctuations $\mathcal{C}_A^{(2)}$ only at sufficiently large length-scales $\ell_A \sim L$, whereas for $\ell_A \ll L$ the behavior is identical to that of the maximally mixed state, Eq.~\eqref{eq:CA2-T-max-mixed}. For arbitrary relation between $\ell_A$ and $L$, in the volume-law regime, we obtain
\begin{equation}
\mathcal{C}_{A}^{(2)}(T)\approx\frac{g\ell_{A}(L-\ell_{A})L^{d-2}}{vT}=\frac{g}{vT}\frac{\left\Vert A\right\Vert \left\Vert B\right\Vert }{\left\Vert A+B\right\Vert },\quad\ell_{A,B}\gg vT,
\label{eq:CA2-max-entangled-pure}
\end{equation}
where the subsystem $B$ is complementary to the subsystem $A$. Owing to a fixed charge of the whole system $A+B$, the variance satisfies $\mathcal{C}_{A}^{(2)}(T)=\mathcal{C}_{B}^{(2)}(T)$, i.e., yields the same result upon replacement $\ell_A \mapsto L - \ell_A$, and, as a consequence, vanishes for $\ell_A = L$.

The diffusive-approximation results for the dynamics of the correlation function
$\mathcal{C}(\boldsymbol{q},T)$ for a maximally mixed initial state, Eq.~\eqref{eq:CqAbsorbing},
and a random bitstring initial state,  Eq.~\eqref{eq:CqReflecting}, are illustrated in Fig.~ \ref{fig:Cq}(a). It is seen that, with increasing $T$, the correlation function approaches the steady-state limit $\mathcal{C}(\boldsymbol{q})=gq$. The behavior of the variance $\mathcal{C}_{A}^{(2)}(T)$ of charge fluctuations for the same initial states is presented in Fig.~\ref{fig:CA2}(a,c). This figure demonstrates how the logarithmic scaling $\mathcal{C}_A^{(2)} = (2g/\pi) L^{d-1} \ln (\ell_A / \ell_0)$
extends to larger length scales as time increases. We will return to a comparison of these results with those of numerical simulations (shown in panel (b) of Fig.~\ref{fig:Cq} and panels (b,d) of Fig.~\ref{fig:CA2}) in Sec.~\ref{sec:diffusive-numerics}.

\subsection{Numerical results}
\label{sec:diffusive-numerics}

To verify and supplement our analytical predictions, we have performed numerical simulations of stochastic projective monitoring of a $d=1$ system, Eq.~\eqref{eq:H}, of size $L = 2000$ with $J = 1$ for a sufficiently small measurement rate $\gamma=0.1$. This value of $\gamma$ is found to belong to the optimal range for simulating the diffusive regime, cf. Ref.~\cite{Poboiko2023a}. When $\gamma$ is taken considerably smaller, the mean free path becomes considerably larger, implying larger corrections due to a vicinity of a crossover to the ballistic regime. On the other hand, taking $\gamma$ substantially larger leads to an increase of weak-localization (or even strong-localization) effects on length scales set by $L$ and $T$. 

\begin{figure}[ht]
    \centering
 \includegraphics[width=\columnwidth]{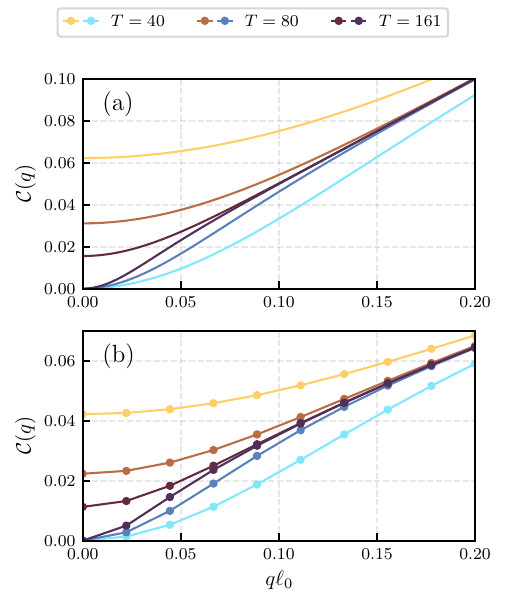}
    \caption{Evolution of the density correlation function $\mathcal{C}(q)$ with time $T$ for a monitored system with a maximally mixed initial state (three top curves in each of the panels) 
    and a random bitstring initial state (three bottom curves in each of the panels). The times $T$ are shown in the legend.
    (a) Analytical results in the diffusive approximation, Eqs.~\eqref{eq:CqAbsorbing} and \eqref{eq:CqReflecting} respectively, with the coupling $g$ and the mean-free path $\ell_0$
corresponding to the parameters of numerical simulations. (b) Numerical results for a $d=1$ system of size $L = 2000$ with the Hamiltonian \eqref{eq:H}, hopping constant $J = 1$ and measurement rate $\gamma = 0.1$. Numerical results for the volume-law pure initial state are indistinguishable from the maximally mixed state, except for the $q=0$ value, $\mathcal{C}(q=0) = 0$, and are not shown here.}
    \label{fig:Cq}
\end{figure}

\begin{figure*}[ht]
    \centering
    \includegraphics[width=\textwidth]{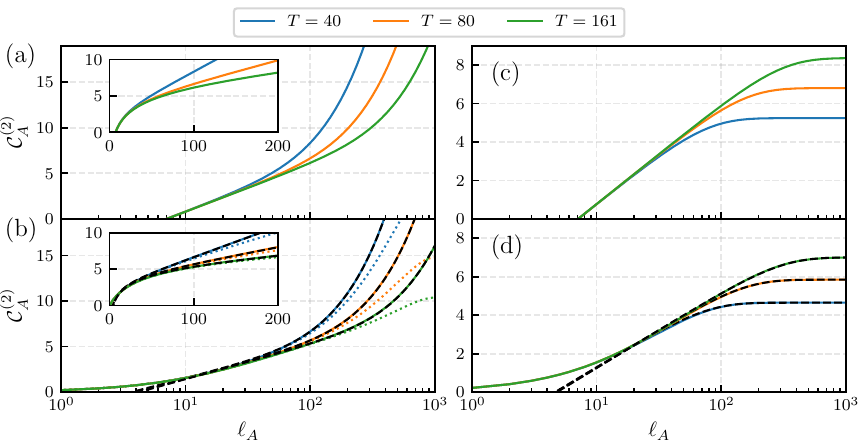}
    \caption{Evolution of the variance of charge fluctuation in a subsystem, $\mathcal{C}_A^{(2)}$,  with time $T$ (shown in the legend). In all the panels, the variance $\mathcal{C}_A^{(2)}$ is shown as a function of the subsystem size $\ell_A$. Top panels: analytical, diffusive-approximation results for a maximally mixed initial state [Eq.~\eqref{eq:CA2-T-max-mixed}, panel (a)] and maximally disentangled initial state [Eq.~\eqref{eq:CA2-disentangled}, panel (c)], with parameters corresponding to those used in numerical simulations. Bottom panels: numerical results for a $d=1$ system of size $L = 2000$ with the Hamiltonian \eqref{eq:H}, hopping constant $J = 1$ and measurement rate $\gamma = 0.1$, for a
    maximally mixed initial state [panel (b), full lines], maximally entangled pure initial state [panel (b), dotted lines], and maximally disentangled initial state [panel (d), full lines]. Dashed lines in panels (b) and (d) are best fits to 
Eqs.~\eqref{eq:CA2-T-max-mixed} and \eqref{eq:CA2-disentangled} with fitting parameters $g$ and $\ell_0$. The values of $g$ obtained from the fits are 
$\sim 20 - 30\%$ below the diffusive-approximation analytical values used in panels (a,c). Insets in the panels (a) and (b) present the same data as in main plots but shown on a linear scale for $x$ axis, which highlights the volume-law behavior at $\ell_A > vT$.
    }
    \label{fig:CA2}
\end{figure*}

We have performed simulations for three types of initial conditions, as discussed in Sec.~\ref{sec:initial_conditions}. A volume-law pure state was implemented as a random bitstring in momentum representation. We have focused on the evolution of the density correlation function $\mathcal{C}(q)$ (Fig.~\ref{fig:Cq}) and of the variance of charge fluctuations in a subsystem $\mathcal{C}_A^{(2)}$ (Fig.~\ref{fig:CA2}) with the time $T$. 

The results for the density correlation function $\mathcal{C}(q)$ for a maximally mixed initial state and a random bitstring initial state are shown in
Fig.~\ref{fig:Cq}(b) for three values of time $T$. For a volume-law pure initial state, we obtain exactly the same results (not shown) as for a maximally mixed initial state, with the only difference that ${\cal C}(\boldsymbol{q}=0) = 0$ instead of $\mathcal{C}(\boldsymbol{q}=0)=g/v T$, in full agreement with the analytical prediction. 

For both types of initial states, the results in Fig.~ \ref{fig:Cq}(b) clearly show a convergence to the steady-state limiting curve. The observed behavior is fully consistent with the analytical predictions in Fig.~\ref{fig:Cq}(a).
Here we used parameters of the NLSM that correspond to the considered microscopic model, $\ell_0 = J / (\gamma \sqrt{2})$, $g = J / (2 \sqrt{2} \gamma)$, and $v = \sqrt{2} J$, see Ref.~\cite{Poboiko2023a}.
Deviations are explained by the fact that, while the numerical simulations are exact, the analytical results are obtained within the diffusive approximation that discards ballistic and localization corrections~\footnote{See Fig.~3 of Ref.~\cite{Poboiko2023a}, which shows the reduction of the effective diffusion constant $g(q) = \mathcal{C}(q) / |q|$ due to ballistic and localization effects in the relevant range of $q$ for various values of $\gamma$, including  $\gamma = 0.1$ used in numerical study in Sec.~\ref{sec:diffusive-numerics} of the present work.}.

In Fig.~\ref{fig:CA2}, we show 
numerical results for the variance 
of charge fluctuation in a subsystem, $\mathcal{C}_A^{(2)}$, for different initial states and compare them to the analytical predictions of Sec.~\ref{sec:diffusive-analytics}.
The numerical results for a maximally mixed and a maximally disentangled initial states are shown by solid lines in panels (b) and (d), respectively. In panels (a) and (c), we present the corresponding analytical results of the diffusive approximation, Eqs.~(\ref{eq:CA2-T-max-mixed}) and (\ref{eq:CA2-disentangled}), with NLSM parameters corresponding to those of the microscopic model. The numerical results excellently demonstrate the analytically predicted behavior. The logarithmic behavior of the variance  holds at $\ell_A \lesssim vT$, thus gradually developing with increasing $T$. At 
$\ell_A \gtrsim vT$, it crosses over into the volume-law behavior for maximally mixed initial state or, respectively, into the area-law behavior (saturation) for maximally disentangled initial state. A closer inspection of the figure reveals some deviations between the values of the variance in analytical and numerical panels. These deviations have the same origin as those discussed above in connection with  Fig.~\ref{fig:Cq}, i.e., they can be attributed to ballistic and weak-localization corrections which are discarded in the diffusive-approximation analytical treatment. As shown by black dashed lines in Fig.~\ref{fig:CA2}(b,d), a nearly perfect fit of the numerical data is obtained if we use the analytical formulas (\ref{eq:CA2-T-max-mixed}) and (\ref{eq:CA2-disentangled}) but consider $g$ and $\ell_0$ as fit parameters. 
The values of $g$ obtained from the fits are 
$\sim 20 - 30\%$ below the diffusive-approximation analytical values (with a weak $T$ dependence), which is the expected manifestation of the corrections discussed above. 

In Fig.~\ref{fig:CA2}(b), we also show by dotted lines numerical results for the case of a volume-law pure-state initial condition. As predicted analytically, they closely follow the data for a maximally mixed initial state for $\ell_A \ll L$. As $\ell_A$ becomes comparable to $L/2$, the data for a maximally entangled pure state begin deviating down in agreement with Eq.~\eqref{eq:CA2-max-entangled-pure}. 
We recall that we considered a specific implementation of a volume-law pure state as a random bitstring in momentum space. The agreement with analytical results supports the expectation that these results hold for a generic volume-law pure state.

\section{Measurement-induced transition via quantum dynamics}
\label{sec:MIPT}

\subsection{Analytical considerations: Finite-size scaling}
\label{sec:MIPT-analytics}

We now turn to the dynamics at larger time scales $T \gtrsim T^\ast$, in the regime where quasi-one-dimensional localization in time direction becomes operative. 
In terms of the NLSM, conductance between the $t=0$ and $t=-T$ boundaries becomes exponentially small 
$\sim\exp\left(-T/T^{\ast}\right)$ in this regime. This implies that correlation functions of interest are nearly equal to their $T\to \infty$ limiting values, with exponentially small corrections $\sim\exp\left(-T/T^{\ast}\right)$. 

We will focus here on systems in spatial dimensionality $d>1$ that exhibit a MIPT at some critical measurement rate $\gamma_c$ and use the long-time dynamics
to detect this transition. This is possible since the quasi-one-dimensional localization length $T^\ast(L,\gamma)$ scales differently with $L$ depending on whether the $(d+1)$-dimensional bulk is delocalized, critical, or localized (i.e., whether $\gamma < \gamma_c$, $\gamma=\gamma_c$, or $\gamma > \gamma_c$). We will discuss this important point in detail below.

For the purpose of determining $T^\ast(L,\gamma)$, it is useful to consider the variance $\mathcal{C}^{(2)}$ of the total charge in the system for a system that is initially in a maximally mixed state. As discussed in Sec.~\ref{sec:initial_conditions}, the charge variance is given by the NLSM conductance $G_{S,S_0}$ between the whole $t=0$ surface and the whole $t=-T$ surface, thus decaying to zero as $\sim\exp\left(-T/T^{\ast}\right)$ at large $T$. The rate of this exponential decay is thus a convenient way to find $T^\ast(L,\gamma)$ in numerical simulations, as will be done below. 

The process of exponential decay of $\mathcal{C}^{(2)}$ has been termed ``charge sharpening'' 
\cite{Agrawal2022,Barratt2022} in the context of monitored quantum circuits. 
Importantly, for the free-fermion class of problems that we consider, the charge variance is directly linked to the entropy, see Eq.~\eqref{eq:S-C}. 
Thus, the entropy of the whole system will decay exponentially at long times $T$ with exactly the same rate $T^\ast(L,\gamma)$. This means that the system will approach a pure state, the process known as  ``purification''
\cite{Gullans2020a}. Therefore, the time scale $T^\ast(L,\gamma)$ can be equivalently called charge sharpening time or purification time in the present context, and we will use these terms interchangeably below. 

The analogy to the Anderson-localization problem is very instructive for understanding of the scaling of $T^\ast(L,\gamma)$.
When the bulk is diffusive, i.e., $\gamma < \gamma_c$, we are in a situation of a quasi-one-dimensional localization problem with $N \sim L^d$ transverse channels and a good mixing between them.  In this case, the localization length is proportional to the number of channels,  i.e., to the $d$-dimensional cross-section of the system, $T^{\ast}(L)\propto N\propto L^{d}$. To obtain this result on a more technical level, we note that, in this regime, the 
NLSM theory \eqref{eq:NLSM} can be approximately restricted to field configurations
$\hat{U}(\boldsymbol{x}, t)$ that depend on the time coordinate $t$
 only, i.e., do not depend on $\boldsymbol{x}$.  The action \eqref{eq:NLSM} then takes the form of a one-dimensional (in time direction) NLSM with the factor $\sim g L^d$ in front. This factor determines the localization length $T^\ast$ in time direction. We also note that the NLSM conductance $G_{S,S_0}$ between the $t=0$ and $t=-T$ boundaries, which is given by Eq.~\eqref{eq:CT:diffusive} at not too long times, becomes of order unity at $T \sim T^\ast$, which is a manifestation of a crossover to the quasi-one-dimensional exponential localization at this time scale.

On the opposite side of the transition, $\gamma > \gamma_c$, where the bulk is localized, the mixing between channels only happens for system sizes smaller than the localization length, $L \lesssim \xi_{\text{loc}}$. A larger system, with $L \gg \xi_{\text{loc}}$,  essentially decouples into independent systems of size  $\xi_{\text{loc}}$. As a result, the quasi-one-dimensional localization length (or, equivalently, purification time) $T^{\ast}$ becomes 
$T^{\ast}\sim\xi_{\text{loc}} /v$ and thus
independent of the system size $L$. 

Finally, at the transition point, $\gamma=\gamma_c$, the system is scale-invariant. The only length scale characterizing such a critical system is the system size $L$ itself, and only it can determine the purification time scale. Thus, $v T^{\ast}(L)\sim L$ at the transition point. In the context of finite-size-scaling analysis of Anderson transitions, the dimensionless constant that is a counterpart of our ratio $v T^{\ast}(L) / L$ at the transition point is conventionally denoted by $\Lambda_c \equiv \Gamma_c^{-1}$ \cite{Kramer2010finite-size,Slevin1997Anderson,Slevin2014critical}.

Summarizing the results of this analysis, we have for the scaling of
$T^{\ast}$ with the system size $L$
(at sufficiently large $L$)
\begin{equation}
\label{eq:TastScaling}
\frac{v T^{\ast}(L, \gamma)}{L} \propto\begin{cases}
L^{d-1}, & \gamma<\gamma_{c},\\
\const, & \gamma=\gamma_{c},\\
L^{-1}, & \gamma>\gamma_{c}.
\end{cases}
\end{equation}
In the vicinity of the transition point, one can further specify the behavior of $T^{\ast}(L, \gamma)$ by using the following scaling considerations. The system near criticality is characterized by a length scale---the localization length $\xi_{\rm loc}$ for $\gamma > \gamma_c$ or the correlation length $\xi_{\rm corr}$ for $\gamma < \gamma_c$---which diverges at criticality as $\xi_{\rm loc}, \ \xi_{\rm corr} \sim |\gamma - \gamma_c|^{-\nu}$. Since microscopic length scales should not be important near criticality, the  dimensionless ratio of the two lengths $v T^{\ast}(L,\gamma)/L$ should depend on $L$ and $\gamma$ via the dimensionless ratio
$L/\xi_{\rm loc}$ or $L/\xi_{\rm corr}$:
\begin{equation}
\label{eq:T-ast-scaling-2}
\frac{v T^{\ast}(L, \gamma)}{L} =
\begin{cases}
{\cal F}_<(L / \xi_{\rm corr}) \,, & \gamma < \gamma_c \,, \\
{\cal F}_>(L / \xi_{\rm loc}) \,, & \gamma > \gamma_c.
\end{cases}
\end{equation}
where  ${\cal F}_<(x) \sim x^{d-1}$ and
${\cal F}_>(x) \sim x^{-1}$
for $x \gg 1$,  and ${\cal F}_<(0) = {\cal F}_>(0)$ is a non-zero constant.
Taking into account the power-law scaling of the localization and correlation lengths near the transition point, this can be rewritten as
\begin{equation}
\label{T-ast-finite-size-scaling}
\frac{v T^{\ast}(L, \gamma)}{L}
=\Psi\left(L^{1/\nu} (\gamma - \gamma_c)\right).
\end{equation}
Evaluating numerically $T^\ast(L,\gamma)$ and fitting the data to the scaling form \eqref{T-ast-finite-size-scaling} allows one to determine the transition point $\gamma_c$ and the associated critical exponent $\nu$.
In the context of Anderson-localization transitions, an analogous finite-size-scaling approach was introduced in Refs.
~\cite{MacKinnon1981one-parameter,Pichard1981finite-size}
and later used in many works to study Anderson transitions of systems of various symmetry classes and spatial dimensionalities, see, e.g.,
\cite{Slevin1997Anderson,
Kramer2010finite-size,
Slevin2014critical} and references therein. 

In Appendix \ref{app:NLSM-T-star-scaling}, we present details of an analysis in the framework of the NLSM that supports the scaling predictions \eqref{eq:T-ast-scaling-2},
\eqref{T-ast-finite-size-scaling}.
For completeness, we also use there the NLSM approach to find the behavior of $T^*(L,\gamma)$ in the case $d=1$, when there is no MIPT but there is a crossover between weak-localization and strong-localization regimes when $L$ increases. 

 We implement the dynamical approach described above in
Sec.~\ref{sec:MIPT-numerics} where we use it to study the MIPT in $d=2$ dimensions. Let us emphasize that this is an  alternative approach to the one employed in Ref.~\cite{Poboiko2023b}, where the position of the transition and the critical exponent were  estimated by considering the \emph{steady-state} behavior of charge covariance between different regions.
It is important to verify consistency between the results of the two approaches (steady-state and dynamical), which will be done below.

\subsection{Numerical study of the MIPT in \texorpdfstring{$d=2$}{d=2} dimensions}
\label{sec:MIPT-numerics}

To study the measurement-induced transition by means of quantum dynamics, we have performed numerical simulations of monitored dynamics of a two-dimensional system with the Hamiltonian \eqref{eq:H} with $J = 1$
on a square lattice of size $L \times L$ with $20 \leq L \leq 52$ subjected to local projective density measurements with the rates $2.0 \leq \gamma \leq 3.7$.
This is exactly the model of Ref.~\cite{Poboiko2023b}, so that a direct comparison will be possible.  

Starting from a maximally mixed initial state, we have calculated the dynamics of variance of charge $\mathcal{C}^{(2)}(T)$ in the whole system as a function of time.
After averaging the logarithm of $\mathcal{C}^{(2)}(T)$
over different quantum trajectories for fixed parameters $(L,\gamma)$, we obtained the typical value 
\begin{equation}
\mathcal{C}^{(2,\text{typ})}(T)\equiv\exp\left(\overline{\ln\mathcal{C}^{(2)}(T)}\right).
\label{eq:C-2-typ-def}
\end{equation}
Its large-time asymptotic behavior was then fitted with an exponential 
\begin{equation}
\mathcal{C}^{(2,\text{typ})}(T)\sim\exp\left(-T/T^{\ast}\right).
\label{eq:C-2-typ-exp}
\end{equation}
In this way, we determined the 
purification (or, equivalently, charge-sharpening) time scale $T^\ast(L, \gamma)$, which plays the role of the time-direction localization length  within the NLSM formalism.
Details of the fitting procedure are presented in Appendix~\ref{sec:app:purification}. We note that the typical value 
$\mathcal{C}^{(2,\text{typ})}(T)$ 
is used here rather than the average one, $\overline{\mathcal{C}^{(2)}(T)}$. The reason is that the typical value, which is directly related to the Lyapunov exponent, has a self-averaging property favorable for its numerical evaluation, while the average value is governed in this regime by exponentially rare events, which makes its numerical evaluation much more difficult. 
This is in analogy with the properties of Anderson localization in quasi-one-dimensional systems and, in particular, with the use of the Lyapunov exponent in finite-size-scaling studies of Anderson transitions \cite{Kramer2010finite-size,Slevin1997Anderson,Slevin2014critical}.

\begin{figure}
    \centering
    \includegraphics[width=\columnwidth]{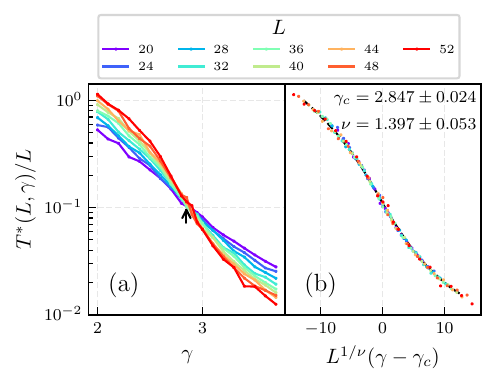}
    \caption{Dynamical scaling analysis of MIPT in a $d=2$ monitored system with $J = 1$ (i.e., $v = \sqrt{2}$) (a)
Ratio $T^\ast(L,\gamma)/L$ of the purification time scale $T^\ast(L,\gamma)$ to the system size $L$, as a function of  measurement rate $\gamma$ for system sizes from $L=20$ to $L=52$ (see legend).  
The  crossing point marked with an arrow provides the position of the  transition point $\gamma_c$. (b) Best single-parameter collapse according to Eq.~\eqref{eq:TastScaling}, which allows us to determine the critical measurement rate $\gamma_c$ and the correlation-length critical exponent $\nu$ (the values shown in the figure) with a good accuracy. Black dashed line marks the scaling function $\Psi(x)$ as a guide to the eye.}
    \label{fig:T}
\end{figure}

The numerically obtained results for $T^\ast(L,\gamma)$ are shown in Fig.~\ref{fig:T}(a). In this figure, the ratio $T^{\ast}(L,\gamma) / L$ is plotted as a function of $\gamma$, with curves of different colors representing different system sizes (see legend). A clear crossing point is observed, which yields the position of the phase transition, $\gamma_c \approx 2.85$. To determine the critical exponent $\nu$,  we have performed the single-parameter collapse of the data according to 
Eq.~\ref{T-ast-finite-size-scaling}, which is shown in Fig.~\ref{fig:T}(b).
The values of $\gamma_c$ and $\nu$ were obtained from the condition of best collapse, and the corresponding error bars were estimated by the subsampling (bootstrap) procedure, see  Appendix \ref{sec:app:collapse} for detail. The obtained values of the critical measurement rate and the correlation-length exponent are 
$\gamma_c = 2.847 \pm 0.024$ and $\nu = 1.397 \pm 0.053$. 

We can now compare the obtained results for $\gamma_c$ and $\nu$ with those found (for the same $d=2$ model) in Ref.~\cite{Poboiko2023b} 
and subsequently in Ref.~\cite{fan2025entanglement}
by studying the scaling of particle-number covariance in the steady state ($T=\infty$).  
The results of Ref.~\cite{Poboiko2023b} are
$\gamma_c = 2.93 \pm 0.09$ and
$\nu = 1.38 \pm 0.09$, and 
Ref.~\cite{fan2025entanglement} obtained, in our notations,
$\gamma_c = 2.86 \pm 0.01$ and
$\nu = 1.31 \pm 0.08$. 
(Here, we have taken into account that $\gamma$ as defined  in Ref.~\cite{fan2025entanglement} is twice larger than the one used in our paper.)  Thus, the values of $\gamma_c$ and $\nu$ obtained via the dynamical approach in the present work are in full agreement (within the error bars) with the values obtained previously from the steady-state analysis. This confirms the consistency of our analytical understanding of the quantum dynamics of monitored free-fermion systems around the MIPT presented in Sec.~\ref{sec:MIPT-analytics}.

\section{Summary and outlook}
\label{sec:summary}

Summarizing, we have studied, by combining analytical arguments and numerical simulations, the dynamical evolution of quantum correlations in monitored free-fermion many-body systems. In the considered setting, the system evolves for a time $T$, starting from a certain initial state, with the evolution governed by a combination of unitary Hamiltonian dynamics and local density measurements. After this, quantum correlations are studied. The analytical approach to the problem is based on a mapping to a NLSM field theory. For a finite evolution time $T$, it should be complemented by boundary conditions, which are analyzed in Sec.~\ref{sec:initial_conditions} for several important classes of initial states (maximally mixed state, random area-law pure state, and random volume-law pure state). 

We further used this general framework in Sec.~\ref{sec:diffusive} to study how quantum correlations gradually ``propagate'' to smaller wave vectors (i.e, larger length scales) with increasing time $T$ in the diffusive regime of not too long times.
We have supported the analytical predictions by numerical simulations of a $d=1$ monitored system with various types of initial states. 

Finally, our focus in Sec.~\ref{sec:MIPT} was on the long-time dynamics, when the system approaches exponentially the steady-state ($T = \infty$) regime, with a characteristic time $T^*$. The geometry of the system in $d+1$ space-time dimensions becomes quasi-one-dimensional in this situation. The time $T^*$, which is simultaneously the purification and the charge-sharpening time, is a counterpart  of the quasi-one-dimensional Anderson-localization length for a $(d+1)$-dimensional disordered system. 
After analyzing finite-size-scaling properties of $T^*(L,\gamma)$ in Sec.~\ref{sec:MIPT-analytics}, we have used this approach to study in
Sec.~\ref{sec:MIPT-numerics} the 
measurement-induced transition in a $d=2$ model. Our numerical results 
for $T^*(L,\gamma)$ satisfy very well the scaling predictions, which has allowed us to extract numerical values of the transition point $\gamma_c$ and the critical exponent $\nu$. The results are in a very good agreement with the values found earlier by a complementary approach based on investigation of steady-state ($T=\infty$) properties. 

Before closing the paper, we discuss prospective directions of study of quantum dynamics and MIPTs in monitored systems that this work is expected to promote. 

\begin{itemize}

\item[(i)]
In Sec.~\ref{sec:MIPT}, we considered the {\it typical} memory-loss time $T^*$,
defined by Eqs.~\eqref{eq:C-2-typ-def} and \eqref{eq:C-2-typ-exp},
which corresponds to the inverse Lyapunov exponent of disordered systems. While $T^*$ is sharply defined in the $T\to\infty$ limit, it exhibits fluctuations at finite $T$, which lead to strong fluctuations of 
$\mathcal{C}^{(2)}(T)$ and related observables around the critical point. 
These fluctuations over the ensemble of quantum trajectories are dynamical counterparts of mesoscopic fluctuations recently studied around the same MIPT in Ref.~\cite{poboiko2025mesoscopic}. 
In particular, they are expected to exhibit multifractal properties at the transition point. We note that, for Anderson transitions in two-dimensional systems,
a relation between correlation-function fluctuations in quasi-one-dimensional geometry and multifractality was predicted by using conformal-mapping arguments and supported by numerical simulations, see Ref.~\cite{Janssen1998statistics}, Sec. 9.2. Investigation of mesoscopic fluctuations of dynamical observables at and around the MIPT is a very interesting prospect for future research.

\item[(ii)]
While our numerics used a model with projective measurements, the NLSM universality suggests that the properties studied in this paper should be the same (after a proper identification of the coupling $g$) for other types of (local) monitoring. The limit opposite to projective measurements is that of continuous monitoring defined via a stochastic Schroedinger equation (also known as quantum-state-diffusion protocol). 
The recent numerical work \cite{fan2025entanglement} confirmed, by studying steady-state properties,  that MIPTs in both models are characterized by the same exponent $\nu$, in agreement with the expected universality. It would be interesting to extend the analysis of the dynamics around the transition on the continuous-measurement model.

\item[(iii)]
Extension of our dynamical analysis onto systems and MIPTs of other universality classes (which includes spatial dimensionality, symmetry class, and topology) represents a broad, promising direction for future research. A challenging goal here is to reach a full understanding of MIPTs of various universality classes and to compare them to Anderson transitions \cite{evers08} described by analogous NLSMs but with a different replica limit. In this context, it is worth mentioning that quantum dynamics in  monitored Majorana systems was discussed in several recent works \cite{Fava2023,XiaoKawabata2024,HisanoriFuji2025}.

\item[(iv)]
Determination of critical exponents of MIPT transitions with high precision would be of great interest, in particular in the context of distinguishing different universality classes. 
To achieve higher precision, it may be important to include sub-leading corrections to scaling, as was done in the context of Anderson transitions \cite{Slevin1999,Slevin2009,Slevin2014critical}. Optimizing models and computational techniques, see, e.g., Ref.~\cite{fan2025entanglement}, would be also beneficial for this purpose.

\item[(v)]
Measurement protocols that make the system non-uniform in the space-time open appealing research prospects.
For example, one can measure with a rate $\gamma_1$ in one part of the space-time and with a rate $\gamma_2$ in the remaining part, thus creating space-time domain walls of any desired geometry. This may be particularly interesting when $\gamma_1$ and $\gamma_2$ are located on opposite sides of a MIPT. See also Ref.~\cite{Pan2025}, which studied a related setup in a different universality class.

In a similar spirit, one can consider non-uniform initial boundary conditions (e.g., different types of states in two halves of the system). 

\item[(vi)]
A recent work demonstrated emergence of superdiffusive scaling of quantum correlations for a free-fermion model with each measurement operator corresponding to the density on a superposition of adjacent sites
\cite{Poboiko2025measurement-induced}.
Observables studied in 
Ref.~\cite{Poboiko2025measurement-induced} were characteristics of the steady state. An important question is how the correlations develop in time in this unconventional class of models. 

\item[(vii)]
Recent works  
\cite{Poboiko2025,guo2025field}
explored the effect of weak interaction in the model of monitored fermions. It was found that the interaction reduces the NLSM symmetry, splitting it into two sectors---one responsible for charge and another one for entanglement. This dramatically affects the phase diagram, leading to emergence of two MIPTs for $d=1$ (where there was no transition at all without interaction) and splitting the MIPT into two transitions for $d >1$. It would be very interesting to explore how this physics manifests itself in dynamical properties of the type of those studied in the present paper. In particular, this will include splitting of purification and charge-sharpening behavior when the interaction is switched on.

\item[(viii)]
Another possible source of non-Gaussianity is to consider a non-Gaussian initial state while keeping the non-interacting (Gaussian) character of evolution. An interesting example of such state is a maximally mixed state in a subspace with a fixed charge. One can then study whether (and how) the Gaussianity of the state is restored in the course of the dynamical evolution.

\end{itemize}

\acknowledgments

We thank Igor Gornyi, Hideaki Obuse, and Tomi Ohtsuki for useful discussions.
We also acknowledge support by the Deutsche Forschungsgemeinschaft (DFG, German Research Foundation) -- 553096561.

\section*{Data availability}

The numerical data and source code that support the findings of this study are openly available in \cite{Data}.

\appendix

\section{Boundary conditions for maximally mixed initial state}
\label{sec:app:boundary}

In this Appendix we provide a derivation of the NLSM boundary conditions for the maximally mixed initial state as described by the many-body density matrix $\hat{\rho}(t_{i})=\hat{\rho}_0=2^{-L^{d}}\hat{\mathbb{I}}$, following the fermionic path integral approach of Ref.~\cite{Poboiko2025}.
For a given quantum trajectory $\mathcal{T}$, the non-normalized density matrix at time $t_f$ can be expressed as $\hat{D}_{{\cal T}}=\hat{S}_{{\cal T}}\hat{\rho}_{0}\hat{S}_{{\cal T}}^{\dagger}$, where $\hat{S}_{\mathcal{T}}$ is a full non-unitary evolution operator, which includes both unitary Hamiltonian evolution and Krauss operators associated with measurements.
In the Keldysh fermionic path integral formalism, the coherent state representation of the density matrix can be expressed as:
\begin{multline}
\hat{D}_{{\cal T}}=\int{\cal D}\bar{\psi}_{i}^{\pm}{\cal D}\psi_{i}^{\pm}\exp\left(-\bar{\psi}_{i}^{+}\psi_{i}^{+}-\bar{\psi}_{i}^{-}\psi_{i}^{-}\right)\\
\times\hat{S}_{{\cal T}}\left|\psi_{i}^{+}\right>\left<\psi_{i}^{+}\right|\hat{\rho}_{0}\left|\psi_{i}^{-}\right>\left<\psi_{i}^{-}\right|\hat{S}_{{\cal T}}^{\dagger},
\end{multline}
where subscript $i$ indicates that fermionic fields correspond to the initial time $t = t_i$.
The matrix element of initial density matrix then trivially yields $\left<\psi_{i}^{+}\right|\hat{\rho}_{0}\left|\psi_{i}^{-}\right>=2^{-L^{d}}\exp\left(\bar{\psi}_{i}^{+}\psi_{i}^{-}\right)$, while evolution operators yield the $\pm$ parts of the Keldysh action. Following standard derivation of the Keldysh path integral representation, see e.g. Ref.~\cite{KamenevLevchenko}, we arrive at:
\begin{multline}
\hat{D}_{{\cal T}}=\int{\cal D}\bar{\psi}^{\pm}(t){\cal D}\psi^{\pm}(t)\exp\left(-\bar{\psi}_{i}^{+}\left(\psi_{i}^{+}-\psi_{i}^{-}\right)\right)\\
\times\left|\psi^{+}_f\right>\exp\left(iS[\psi^{\pm},\psi^{\pm}]-\bar{\psi}_{f}^{-}\psi_{f}^{-}\right)\left<\psi_{f}^{-}\right|.
\end{multline}
In the continuous time representation, the argument of the first exponential factor then implies the initial condition for the fermionic fields $\psi^{+}(t_{i})=\psi^{-}(t_{i})$; the same condition should hold for each individual replica of the fermionic fields when considering $R \to 1$ copies of the non-normalized density matrix. 

The NLSM matrix field $\hat{U}(t)$ describes Goldstone modes associated with unitary rotations $\hat{\mathcal{V}}_{\pm}(t)$ of the $\psi^{\pm}(t)$ fields in the replica space, see Ref.~\cite{Poboiko2025}, as $\hat{U}(t)=\hat{{\cal V}}_{+}(t)\hat{{\cal V}}_{-}^{\dagger}(t) \in \mathrm{SU}(R)\times\mathrm{SU}(R)/\mathrm{SU}(R)$. 
The initial condition for fermionic fields allows only identical rotations $\hat{\mathcal{V}}_{+}(t_{i})=\hat{\mathcal{V}}_{-}(t_{i})$, which finally yields the initial condition $\hat{U}(t_{i})=\hat{\mathbb{I}}$ for the NLSM field.

\section{NLSM representation of density correlation function}
\label{sec:app:NLSM:current}

In this Appendix, we present a derivation of the expression for density correlation function in the NLSM language, Eq.~\eqref{eq:C} of the main text. Our starting point is the charge generating function introduced in Ref.~\cite{Poboiko2025}:
\begin{equation}
Z[\lambda]=\overline{\prod^{R}_{r=1}\left\langle \exp\left(i\int d^{d}\boldsymbol{x}\lambda_{r}(\boldsymbol{x})\hat{n}(\boldsymbol{x})\right)\right\rangle},
\end{equation}
where we restrict ourselves to the sources satisfying $\sum^{R}_{r=1}\lambda_{r}(\boldsymbol{x})=0$. Similarly to the derivation in Appendix~\ref{sec:app:boundary}, counting fields $\lambda_r(\boldsymbol{x})$ can be absorbed in the boundary condition for fermionic fields at time $t = t_f$ in a form $\psi^{-}_{r}(\boldsymbol{x},t_f) = e^{i \lambda_r(\boldsymbol{x})} \psi^{+}_{\boldsymbol{r}}(\boldsymbol{x},t_f)$. The latter then straightforwardly transforms into the boundary condition for the NLSM, $\hat{U}(\boldsymbol{x},t_{f})=\exp\left(i\hat{\lambda}(\boldsymbol{x})\right)$, with the diagonal matrix $\hat{\lambda}(\boldsymbol{x})$:
\begin{equation}
Z[\lambda]=\int_{\hat{U}(\boldsymbol{x},t_{f})=e^{i\hat{\lambda}(\boldsymbol{x})}}{\cal D}\hat{U}\exp\left(-S[\hat{U}]\right).
\end{equation}
The specific choice $\lambda_a(\boldsymbol{x}) = \lambda_1 \delta(\boldsymbol{x} - \boldsymbol{x}_1) + \lambda_2 \delta(\boldsymbol{x} - \boldsymbol{x}_2)$ and $\lambda_b(\boldsymbol{x)} = -\lambda_a(\boldsymbol{x})$ for two arbitrary replicas $a \neq b$ then allows us to express the density correlation function as follows:
\begin{equation}
\mathcal{C}(\boldsymbol{x}_{1},\boldsymbol{x}_{2})=-\frac{1}{2}\left.\frac{\partial^{2}\ln Z[\lambda]}{\partial\lambda_{1}\partial\lambda_{2}}\right|_{\lambda_{1,2}=0}.
\end{equation}

In order to calculate $Z[\lambda]$, we utilize $\mathrm{SU}(R)$ symmetry of the action and perform a gauge transformation $\hat{U}(\boldsymbol{x},t)=\hat{U}^{\prime}(\boldsymbol{x},t)\exp\left(i\chi(t)\hat{\lambda}(\boldsymbol{x})\right)$ with a function $\chi(t)$ satisfying
$\chi(t = t_f) = 1$ and sharply dropping to zero on a very short time-scale $\tau \to 0$. Such gauge transformation makes boundary condition trivial $\hat{U}^{\prime}(\boldsymbol{x},t_{f})=\hat{\mathbb{I}}$ and generates additional terms in the action. The only terms surviving in the limit $\tau \to 0$  are generated from the temporal derivatives and read:
\begin{equation}
S[\hat{U}]=S[\hat{U}^{\prime}]+\frac{g}{v}\int d^{d}\boldsymbol{x}dt\left(\dot{\chi}\Tr\left(\hat{{\cal J}}^\prime\hat{\lambda}\right)+\frac{1}{2}\dot{\chi}^{2}\Tr\left(\hat{\lambda}^{2}\right)\right),
\end{equation}
with the Noether current $\hat{\mathcal{J}}^{\prime}=-i\hat{U}^{\prime\dagger}\partial_{t}\hat{U}^{\prime}$. We then expand the generating function to the second order in $\hat{\lambda}$ and integrate by parts over time to get rid of arbitrary function $\chi(t)$, arriving at:
\begin{multline}
\ln Z[\lambda]\approx-\frac{g}{2v}\int d^{d}\boldsymbol{x}_{1,2}\lim_{t_{1,2}\to t_f}\sum_{ab}\lambda_{a}(\boldsymbol{x}_{1})\lambda_{b}(\boldsymbol{x}_{2})\\\times\Bigg(\delta_{ab}\delta(\boldsymbol{x}_{1}-\boldsymbol{x}_{2})\delta(t_{1}-t_{2})-\frac{g}{v}\left\langle {\cal J}_{aa}(\boldsymbol{x}_{1},t_{1}){\cal J}_{bb}(\boldsymbol{x}_{2},t_{2})\right\rangle,
\end{multline}
where we have omitted primes for brevity. For our choice of the sources this yields:
\begin{multline}
\label{eq:app:Cx1x2-NLSM}
\mathcal{C}(\boldsymbol{x}_{1},\boldsymbol{x}_{2})=\lim_{t_{1,2}\to t_f}\Bigg[\frac{g}{v}\delta(\boldsymbol{x}_{1}-\boldsymbol{x}_{2})\delta(t_{1}-t_{2})\\-\frac{g^{2}}{v^{2}}\Bigg(\frac{1}{2}\left\langle {\cal J}_{aa}(\boldsymbol{x}_{1},t_{1}){\cal J}_{aa}(\boldsymbol{x}_{2},t_{2})\right\rangle +\frac{1}{2}\left\langle {\cal J}_{bb}(\boldsymbol{x}_{1},t_{1}){\cal J}_{bb}(\boldsymbol{x}_{2},t_{2})\right\rangle \\-\left\langle {\cal J}_{aa}(\boldsymbol{x}_{1},t_{1}){\cal J}_{bb}(\boldsymbol{x}_{2},t_{2})\right\rangle \Bigg)\Bigg].
\end{multline}
Finally, the global $\mathrm{SU}(R)$ symmetry of the action implies the following structure of the correlation functions in the replica space:
\begin{equation}
\left\langle {\cal J}_{ij}(\boldsymbol{x}_{1}){\cal J}_{kl}(\boldsymbol{x}_{2})\right\rangle =F(\boldsymbol{x}_{1},\boldsymbol{x}_{2})\left(\delta_{il}\delta_{kj}-\frac{1}{R}\delta_{ij}\delta_{kl}\right),
\end{equation}
which allows us to equivalently rewrite the NLSM representation \eqref{eq:app:Cx1x2-NLSM} of the density correlation function  in the form given by Eq.~\eqref{eq:C}.

\section{Scaling of purification time \texorpdfstring{$T^\ast(L,\gamma)$}{T*(L,gamma)} from NLSM renormalization group }
\label{app:NLSM-T-star-scaling}

Here we provide details of analytical arguments based on NLSM renormalization group (RG) for scaling of $T^\ast(L,\gamma)$ around MIPT for $d>1$ spatial dimensions, Sec.~\ref{sec:MIPT-analytics} of the main text.
For completeness, we also use the same framework to obtain the scaling of $T^\ast(L,\gamma)$ for $d=1$. 

The RG flow for the coupling 
$G(\ell) = g(\ell)\ell^{d-1}$ 
(a counterpart of the dimensionless conductance of the Anderson-localization problem) is governed, to the one-loop approximation, by the equation \cite{Poboiko2023b}
\begin{align}
\frac{dG}{d\ln \ell}&\equiv\beta(G)=\epsilon\, G-\frac{1}{4\pi}+O(1/G),
\label{RG-beta}
\end{align}
where $\epsilon = d-1$. 
For $d > 1$, there is a transition, with the critical coupling $G_c$ determined by the equation $\beta(G_c)=0$ and the critical exponent $\nu$ given by $\nu = [\beta'(G_c)]^{-1}$.  For $\epsilon \ll 1$, one finds 
\begin{equation}
G_c = \frac{1}{4\pi\epsilon} + O (\epsilon^0) \,; \qquad 
\nu = \frac{1}{\epsilon} + O (\epsilon^0).
\label{eq:Gc_nu_epsilon}
\end{equation}
For $d=2$ (i.e., $\epsilon=1$), 
equations \eqref{eq:Gc_nu_epsilon} are not parametrically controlled; they  can serve as rough estimates of $G_c$ and $\nu$. 
For $G_0 \equiv G(\ell_0)$ close to $G_c$, one can integrate the RG equation near criticality, obtaining
$G(\ell)-G_c \simeq (G_0-G_c) (\ell/\ell_0)^{1/\nu}$. The length scale $\ell$ at which the difference $|G(\ell)-G_c|$ becomes of order $G_c$ yields the localization length $\xi_{\rm loc}$ for $G_0 < G_c$ or the 
correlation length $\xi_{\rm corr}$ for $G_0 > G_c$:
\begin{equation}
\xi_{\rm loc}, \ \xi_{\rm corr}
\sim \ell_0 \left( \frac{|G_0-G_c|}{G_c} \right)^{-\nu}.
\end{equation}

We use now this formalism to demonstrate the scaling \eqref{eq:T-ast-scaling-2}
of the purification time $T^\ast(L,\gamma)$ on both sides of the MIPT for $d > 1$. Consider first the case $\gamma > \gamma_c$, i.e., $G_0 < G_c$. When RG is performed from the scale $\ell_0$ to $L$, there are two possibilities. If $L \ll \xi_{\rm loc}$, the system will remain close to criticality up to $\ell \sim L$, implying that $G(L) \approx G_c$. Then $v T^* \sim L$ as at the critical point. If  $L \gg \xi_{\rm loc}$, the system enters the strong-localization regime at $\ell \sim \xi_{\rm loc}$, which essentially stops the RG. In this case, $v T^* \approx \xi_{\rm loc}$. These results provide both asymptotics of the function ${\cal F}_>(x)$ in 
Eq.~\eqref{eq:T-ast-scaling-2}.

Now we consider the case $\gamma < \gamma_c$, i.e., $G_0 > G_c$. Again, there are two possibilities for the outcome of running the RG in $d+1$ dimensions from $\ell_0$ to $L$.  If $L \ll \xi_{\rm corr}$, the system remains critical all the way up to the scale $L$, yielding $v T^* \sim L$. If $L \gg \xi_{\rm corr}$,
the system enters the diffusive regime for $\ell > \xi_{\rm corr}$, where 
the scaling is governed by the term $(d-1)G$ on the right-hand-side of Eq.~\eqref{RG-beta} and the loop correction can be neglected, implying that $G(L) \sim (L/\xi_{\rm corr})^{d-1}$. The scale $L$, where the RG in (d+1) dimensions terminates, serves then as a starting point for one-dimensional renormalization in time direction (corresponding effectively to $d=0$). We will then have a diffusive behavior, $G(v T) \sim G(L) L / v T$ as long as $G(v T)$ remains larger than unity. The condition $G(v T) \sim 1$ will mark the entrance into the localized regime and thus the scale $T^*$, yielding
\begin{equation}
v T^* \sim L G(L)  \sim L (L/\xi_{\rm corr})^{d-1} \,.
\label{eq:T-star-scaling-deloc}
\end{equation}
We have thus obtained
both asymptotics of the function ${\cal F}_<(x)$ in 
Eq.~\eqref{eq:T-ast-scaling-2}.

The purification time ($T^*$ in our notations) in a $d=2$ free-fermion model was numerically considered in Ref.~\cite{chahine2023entanglement}, with a conclusion that it scales according to a power law, $T^* \propto L^\alpha$, with an exponent $\alpha$ continuously decreasing from 2 to 0 when the measurement rate $\gamma$ increases. This is clearly in contradiction with our result \eqref{eq:TastScaling}, according to which the asymptotic (large-$L$) exponent $\alpha$ takes only three values: 2 in the delocalized phase ($\gamma < \gamma_c$), 1 at the transition point ($\gamma = \gamma_c$), and 0 in the localized phase ($\gamma > \gamma_c$). The continuously changing exponent $\alpha(\gamma)$ extracted numerically 
in Ref.~\cite{chahine2023entanglement} is in fact a manifestation of finite-size effects described by the scaling functions in Eqs.~\eqref{eq:T-ast-scaling-2} and \eqref{T-ast-finite-size-scaling} and fully supported by our numerical results in Sec.~\ref{sec:MIPT-numerics}. This ``finite-size exponent'' should approach the asymptotic one (with a jump at $\gamma_c$ from $\alpha=2$ through $\alpha=1$ to $\alpha=0$) with increasing $L$.

We analyze now the scaling of $T^*(L,\gamma)$ for $d=1$. In this case, there is no transition in the $L\to \infty$ limit but there is a crossover at finite $L$ between a diffusive regime (with weak-localization correction) and a strong-localization regime, which is reflected in the scaling of the purification time. If $\gamma \gtrsim 1$, we have $\xi_{\rm loc} \lesssim 1$ and $v T^* \sim \xi_{\rm loc}$. If $\gamma$ is small, the system in $d+1 = 2$ space-time dimensions is still localized in the asymptotic limit $L \to \infty$ but the localization length is large,
$\xi_{\rm loc} \sim \ell_0\exp(1/4\pi G_0)$, as follows from the solution of the one-loop RG equation,
\begin{equation}
G(\ell) \simeq G_0 - \frac{1}{4\pi} \ln \frac{\ell}{\ell_0} \,.
\end{equation}
There are thus two situations, depending on the relation between the system size $L$ and the localization length. If $L \ll \xi_{\rm loc}$,  the time $T^*$ is given by $L G(L) / v$, cf. Eq.~\eqref{eq:T-star-scaling-deloc}.
In the opposite case, $L \gg \xi_{\rm loc}$, we get $v T^* \sim \xi_{\rm loc}$. Therefore, we have
\begin{equation}
\frac{v T^*}{L} \sim \left\{
\begin{array}{ll}
\displaystyle G_0 - \frac{1}{4\pi} \ln \frac{L} {\ell_0} \,, \quad & L \ll \xi_{\rm loc}\,, \\[0.35cm]
\displaystyle \frac{\xi_{\rm loc}}{L} \,, \quad &
 L \gg \xi_{\rm loc}\,.
\end{array}
\right.
\label{eq:T-star-1D}
\end{equation}

The authors of Ref.~\cite{Loio2023} studied numerically the purification time in a $d=1$ free-fermion model with U(1) conservation law and interpreted the results in terms of a Berezinskii-Kosterlitz-Thouless transition with a power-law scaling $T^* \propto L^\alpha$ in the delocalized phase. According to the above analysis, this interpretation is not correct:  there is no transition and $T^*$ saturates at $\xi_{\rm loc} / v$ in the large-$L$ limit. We argue that the apparent power-law scaling in the range of system sizes considered in Ref.~\cite{Loio2023} in fact corresponds to the weak localization (and the crossover to strong localization) in Eq.~\eqref{eq:T-star-1D}. Indeed, the numerically fitted exponent $\alpha$ in Ref.~\cite{Loio2023} exhibits a decrease with $L$, in consistency with the saturation $T^* \to \xi_{\rm loc} / v$, and thus $\alpha \to 0$, at $L\to \infty$.

\section{Numerical study around the MIPT: Purification time \texorpdfstring{$T^\ast(L,\gamma)$}{T*(L,gamma)}}
\label{sec:app:purification}

Here we present details on how the purification time $T^\ast(L,\gamma)$ was extracted from numerical simulations of the dynamics of a $d=2$ monitored system around the MPIT, Sec.~\ref{sec:MIPT-numerics}.

\begin{figure}[ht]
    \centering
\includegraphics[width=\columnwidth]{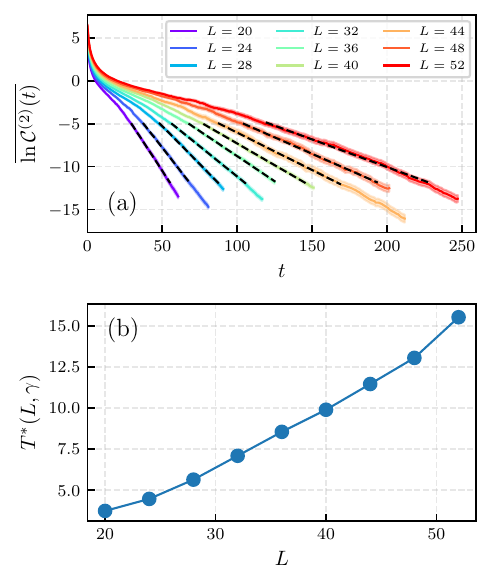}
    \caption{Example of long-time quantum dynamics in a $d=2$ system with $J = 1$ on the delocalized side of the MIPT, $\gamma = 2.6$, for different system sizes $L$.
    (a) Numerical results for time dependence of 
    $\overline{\ln\mathcal{C}^{(2)}(T)}$ characterizing the typical variance of
    charge fluctuations.
    Dashed lines show exponential fits
$\mathcal{C}^{(2,\text{typ})}(T)\sim\exp\left(-T/T^{\ast}\right)$, from which the values of $T^{\ast}(L,\gamma)$ are obtained.       
    (b) Purification time $T^{\ast}(L,\gamma)$ extracted from fitting of the exponential tail in panel (a) as a function of system size.}
    \label{fig:Tfit:2.6}
\end{figure}

Figure \ref{fig:Tfit:2.6}(a) shows the time dependence of $\ln \mathcal{C}^{(2)}(t)$ averaged over different trajectories for $\gamma = 2.6$. The width of each line corresponds to the standard error of the mean value owing to trajectory-to-trajectory fluctuations. We observe the power-law behavior at small times, according to Eq.~\eqref{eq:CT:diffusive}, which then at larger times changes to an exponential decay. The slope of the exponential decay extracted from fits then yields the purification timescale $T^\ast(L,\gamma)$, which is plotted in Fig.~\ref{fig:Tfit:2.6}(b). We observe a faster-than-linear 
increase of $T^\ast(L,\gamma)$
with $L$, which is characteristic for the delocalized phase, $\gamma < \gamma_c$. For our largest $L$, the behavior approaches the $T^\ast(L,\gamma) \propto L^2$ scaling
expected in the large-$L$ limit, see Eq.~\eqref{eq:TastScaling}. A slower than $L^2$ (but faster than $L$) increase of $T^\ast(L,\gamma)$ at intermediate length scales is related to the fact that the value of $\gamma = 2.6$ is quite close to the critical one, implying that the correlation length $\xi_{\rm corr}$ is quite large. When the system size $L$ is comparable to $\xi_{\rm corr}$, one observes a crossover from the critical behavior at $L \ll \xi_{\rm corr}$ to the diffusive behavior at $L \gg \xi_{\rm corr}$, which is described by the scaling function ${\cal F}_<(L/\xi_{\rm corr})$ in Eq.~\eqref{eq:T-ast-scaling-2}.

\begin{figure}[ht]
    \centering
   \includegraphics[width=\columnwidth]{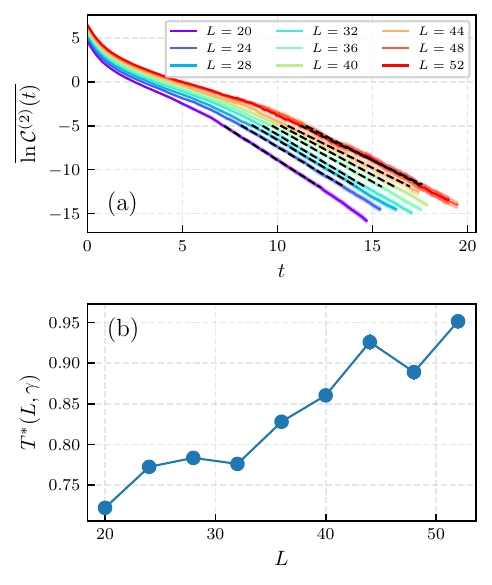}
    \caption{
    Example of long-time quantum dynamics in a $d=2$ system with $J = 1$ on the localized side of the MIPT, $\gamma = 3.5$, for different system sizes $L$.
    (a) Numerical results for time dependence of 
    $\overline{\ln\mathcal{C}^{(2)}(T)}$ characterizing the typical variance of
    charge fluctuations.
    Dashed lines show exponential fits
$\mathcal{C}^{(2,\text{typ})}(T)\sim\exp\left(-T/T^{\ast}\right)$, from which the values of $T^{\ast}(L,\gamma)$ are obtained.       
    (b) Purification time $T^{\ast}(L,\gamma)$ extracted from fitting of the exponential tail in panel (a) as a function of system size.}
    \label{fig:Tfit:3.5}
\end{figure}

In Fig.~\ref{fig:Tfit:3.5}, we provide analogous data for $\gamma = 3.5$. We observe a very different behavior: the extracted purification timescale $T^{\ast}(L,\gamma)$ is nearly $L$-independent. This behavior is characteristic of the localized phase, where $T^{\ast}(L,\gamma)$ is predicted to be $L$-independent in the large-$L$ limit.
The observed residual growth 
(much slower than linear) has essentially the same explanation as the fact that the data in Fig.~\ref{fig:Tfit:2.6}(b) has not reached yet the asymptotic $L^2$ scaling (see previous paragraph). 
The value $\gamma = 3.5$ is quite close to the critical one, implying that the localization length $\xi_{\rm loc}$ is quite large. When the system size $L$ is comparable to $\xi_{\rm loc}$, one observes a crossover from the critical behavior at $L \ll \xi_{corr}$ to the localized behavior at $L \gg \xi_{\rm corr}$, which is described by the scaling function ${\cal F}_>(L/\xi_{\rm corr})$ in Eq.~\eqref{eq:T-ast-scaling-2}.

The numerical procedure described in this Appendix was performed for measurement rates $\gamma$ in the range from 2.0 to 3.7, yielding the data shown in Fig.~\ref{fig:T}(a) of the main text.

\section{Finite-size scaling collapse analysis}
\label{sec:app:collapse}

In this Appendix, we provide technical details on scaling collapse of data for $T^{\ast}(L,\gamma)$ that was carried out in Sec.~\ref{sec:MIPT-numerics} to determine the position of the transition and the correlation-length critical exponent. 

\begin{figure}[ht]
    \centering
\includegraphics[width=\columnwidth]{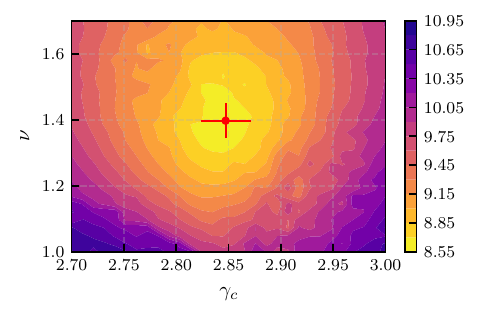}
    \caption{Heatmap of the quality function $\chi^2(\gamma_c,\nu)$ characterizing the scaling collapse, see text for detail. 
    The red dot with error bars marks the values of $\gamma_c$ and $\nu$ determined by the minimum and the width of this function as obtained by the subsampling (bootstrap) procedure.
    }
    \label{fig:collapse-heatmap}
\end{figure}

For given values of the parameters $\gamma_c$ and $\nu$, we perform rescaling of all data points according to Eq.~\eqref{T-ast-finite-size-scaling}, denoting $x = L^{1/\nu} (\gamma - \gamma_c)$ and $y = T^\ast(L,\gamma) / L$, and then analyze how well the single-parameter scaling $y=\Psi(x)$ holds. For this purpose, we approximate the ``master curve'' $\Psi(x)$ with a quadratic spline, whose parameters are fitted by minimizing the $\chi^2$ quality function:
\begin{equation}
\chi^2[\Psi]=\sum_{i}\left(\frac{y_{i}-\Psi(x_{i})}{\delta y_{i}}\right)^{2},
\end{equation}
where $\delta y_i$ denotes the statistical uncertainty of $y_i$. The best fit then yields a numerical characteristic of the collapse quality for given values of the parameters $\gamma_c$ and $\nu$, that we denote as $\chi^2(\gamma_c, \nu)$. The heatmap of the quality function $\chi^2(\gamma_c, \nu)$ obtained in this way is presented on Fig.~\ref{fig:collapse-heatmap}.

Minimization of the $\chi^2(\gamma_c,\nu)$ quality function with respect to its arguments then yields the best estimates for the position of the transition $\gamma_c$ and the critical exponent $\nu$, and the effective width of the minimum yields estimates of the associated error bars. To implement this in a robust way, 
we have resorted to the subsampling procedure \cite{Subsampling1999}, also known as bootstrap. We have randomly chosen $N_s = 500$ subsamples, each containing $\sim 50\%$ of all data points, and estimated the position of the minima for each subsample. 
This yielded the values and the error bars of the transition point and the correlation-length exponent,
$\gamma_c = 2.847 \pm 0.024$ and $\nu = 1.397 \pm 0.053$, which are presented in Sec.~\ref{sec:MIPT-numerics} and are also
shown in Fig.~\ref{fig:collapse-heatmap}. 

\pagebreak

\bibliography{refs.bib}

\end{document}